\documentclass[12pt]{article}
\title{Glueballs,  gluerings and gluestars in the $d=2+1$ SU(N)  gauge theory
}
\author{ Yu.A.Simonov\\
State Research
Center\\Institute of Theoretical and Experimental Physics, \\
Moscow, 117218 Russia}

\newcommand{\beq}{\begin{eqnarray}}
 \newcommand{\eeq}{\end{eqnarray}}
\newcommand{\be}{\begin{equation}}
 \newcommand{\ee}{\end{equation}}

 \def\la{\mathrel{\mathpalette\fun <}}

\def\fun#1#2{\lower3.6pt\vbox{\baselineskip0pt\lineskip.9pt
\ialign{$\mathsurround=0pt#1\hfil ##\hfil$\crcr#2\crcr\sim\crcr}}}

\newcommand{{\SD}}{\rm SD}

\newcommand{\veY}{\mbox{\boldmath${\rm Y}$}}

\newcommand{\ver}{\mbox{\boldmath${\rm r}$}}

\newcommand{\vep}{\mbox{\boldmath${\rm p}$}}

\newcommand{\veR}{\mbox{\boldmath${\rm R}$}}

\newcommand{\vexi}{\mbox{\boldmath${\rm \xi}$}}
\newcommand{\veta}{\mbox{\boldmath${\rm \eta}$}}

\newcommand{\lan}{\langle}
\newcommand{\ran}{\rangle}

\begin{document}
\maketitle
\begin{abstract}
 The $3d$ gluodynamics which governs  the large $T$ quark gluon plasma is studied in the framework of the field correlator method.
 Field correlators and spacial string tension  are derived  through the gluelump Green's
 functions. The glueball spectrum is calculated both in $C=-1$ as well as in $C=+1$ sectors, and
 multigluon bound states in the form of "gluon rings" and "gluon stars" are computed explicitly. Good overall  agreement
  with available lattice data is observed.

\end{abstract}

\section{Introduction}

As  was suggested fifteen years ago \cite{r1,r2}, in QCD  magnetic
fields survive the temperature phase transition and keep their
magnitude almost intact up to $T=1.5 ~T_c$, as was checked by
lattice measurements of field correlators \cite{r3}.

At larger $T, ~T\geq 2T_c$, magnetic fields as measured by the
spacial string tension $\sigma_s (T)$ start to grow quadratically,
$\sigma_s (T) \sim T^2$ \cite{r4,r5}, which signifies the advent
of a new regime, called the dimensional reduction. In this regime
the temporal direction is squeezed, $\Delta t =\frac{1}{T} \to 0$,
while higher Matsubara frequencies are suppressed, so the
effective dynamics is reduced to $3d$ gluodynamics
\cite{r6}-\cite{r9}. The latter is a confining theory and usually
one relies on lattice $3d$ calculations to obtain physical
quantities, such as $\sigma_s (T)$ \cite{r4,r7}.

Instead we suggest here to use  the Field Correlator Method
\cite{r10}, (see \cite{r11} for a review) to study the $3d$
nonperturbative dynamics  and  in particular to compute field
correlators, string tension
 and lowest mass excitations in
$3d$ gluodynamics analytically. Knowing $\sigma_s(T)$ one can
compute other physical quantities of interest, e.g. in \cite{r12}
the $3 d$ screening masses of mesons and of the lowest  glueball
were obtained in good agreement with existing lattice data.

To find  $\sigma_s (T)$ one needs to know the nonperturbative
correlator of magnetic fields $D^H(x)$ \cite{r1,r2}. The latter
can be studied analytically using the new method suggested in
\cite{r13}, where it was expressed in terms of the  gluelump
Green's functions. In case of $4 d$ QCD  the gluelumps were
studied in \cite{r14}, and in the present paper this work is
extended to $3 d$. The confining correlator $D^H(x)$  is expressed
through  the  two-gluon gluelump Green's function,  and for the
latter one has  to use the relativistic 3-body dynamics, which is
well developed in Refs. \cite{r15}-\cite{r16}.  Another part of
our study is the calculation of the gluonic bound states:
glueballs, gluerings and gluestars. We are using the same
Hamiltonian approach as in the $4d$ case \cite{r17},\cite{r18} to
compute masses of these states in terms of the only parameter of
the theory, $\sigma_s (t)$.  The resulting ratios
$M_n/\sqrt{\sigma_s}$ are in  good agreement with existing lattice
data.

 The
plan of the paper is as follows. In section 2 the main formalism
is introduced, and  the 3 body gluelump Green's function is
calculated, together with $\sigma_s(T)$. In section 3  the
glueballs are calculated in the $3d$ gluodynamics, and in section
4 gluerings and gluestars are introduced and the masses are
computed.  In the last section a comparison to lattice and other
approaches is given and a summary and prospectives are outlined.

\section{The large $T$ limit of QCD}

The QCD action $S$ at finite temperature \be S=S_g+S_q;~~ S_g
=\frac{1}{4g^2} \int^\beta_0 dx_4 \int d^2 x (F^a_{\mu\nu})^2,
~~S_q=\int^\beta_0 dx_4 \int d^2x\bar \psi(m-\hat D)
\psi\label{r1}\ee at large $T$ goes over into the effective $3d$
theory \cite{r6}-\cite{r9}, where the leading part of $S_g$
transforms as \be S_g=\frac{1}{g^2T} \int (F_{ik}^a)^2 d^3 x
\equiv\frac{1}{g^2_3 }\int (F^a_{ik})^2 d^3x \label{r2}\ee with
$g^2_3\equiv g^2 T$ is a dimensionful coupling constant of the
$3d$ theory. It is also known
 that quark degrees of freedom decouple while
$A_4$ can be effectively integrated out at large $T$
\cite{r6}-\cite{r9}.

Our aim is the calculation of field correlators in $3d$ theory
using the gluelump Green's function, and to this end  we absorb
$g_3$ in  $F_{ik}$ as $\bar F_{ik} =\frac{1}{g_3}F_{ik}$ and we
define the correlator in the same way as in the  $4d$
case\cite{r10,r11}
$$ \frac{1}{N_c} \lan tr (g_3 \bar F_{ik} (x) \Phi (x,y) g_3 \bar
F_{em}(y) \Phi (y,x)\ran \equiv D_{ik, lm} x,y)=$$ \be
=(\delta_{il} \delta_{km} -\delta_{im} \delta_{kl}) D^H (x-y)
+\frac12 [\partial_i h_l \delta_{km} +perm]
D_1^H(x-y).\label{r3}\ee

The spacial string tension $\sigma_s$ is  expressed through
$D^H(x)$ in the usual way \cite{r1,r2,r19} \be \sigma_s=\frac12
\int D^H(x) d^2x.\label{r4}\ee Note that we  keep index $H$ in
$D^H(x)$, appropriate for $4d$ theory, since $D^H$ coincides with
the high $T$ limit of $4d$ magnetic correlator $D^H_{d=4}$.

To proceed one can use as in \cite{r13} the Background
Perturbation Theory (BPTh) \cite{r20} and write in the  lowest
order of BPTh (cf. Eq. (35) of \cite{r13}) \be D^H_{ik,lm} (x,y)
=-\frac{g^4_3}{2N_c^2} \lan tr_a([a_i,a_k]\hat \Phi(x,y) [a_l,
a_m])\ran\label{r5}\ee where $\hat \Phi$ and $tr_a$ are the
parallel transporter and trace in adjoint representation
respectively.

Writing $[a_i, a_k]= ia_i^aa_k^b f^{abc} T^c$ one can define the
2-gluon gluelump Green's function $G^{(2gl)} (x,y)$ as $$
D^H_{ik,lm} (x,y) = \frac{g^4_3}{2N_c^2} tr_a\lan f^{abc} f^{def}
a_i^a(x) a_k^b (x) T^c\hat \Phi(x,y) T^fa^d_l (y) a^e_m (y)\ran=$$
\be\frac{g^4_3(N^2_c-1)}{2} (\delta_{il}\delta_{km} -\delta_{im}
\delta_{kl}) G^{(2gl)} (x,y)\label{r6}\ee and hence from
(\ref{r3}) one has \be D^H(x,y) =\frac{g^4_3(N^2_c-1)}{2} G^{2gl)}
(x,y).\label{r7}\ee

The two-gluon gluelump Green's function $G^{(2gl)}$ in $4d$ was
studied and numerically evaluated in \cite{r14}, and in what
follows we shall repeat the same calculations for the lowest state
in $3d$.

In the lowest order in $g_3$ the BPTh gives for $G^{2gl)}$ the
following  Fock-Feynman-Schwinger (FFS) path-integral
representation \cite{r21} \be G^{(2gl)} (x,y) =\int^\infty_0 ds_1
\int^\infty_0 ds_2 (Dz^{(1)})_{xy} (Dz^{(2)})_{xy} e^{-K_1-K_2}
\lan W_F\ran\label{r8}\ee where \be W_F = PP_F \exp ig_3\int^y_x
A_i dz^{(1)}_i \exp ig_3\int^y_x A_k
dz_k^{(2)}U_F(x,y)\label{r9}\ee and \be U_F(x,y) =\exp
2ig_3\int^{s_1}_0 d\tau_1 \hat F(z_1(\tau_1))\exp 2ig_3
\int^{s_2}_0 d\tau_2 \hat F(z_2(\tau_2))\label{r10}\ee

As a next step we neglect  the spin interaction of gluons due to
the term $U_F$ in (\ref{r9}), which was studied in \cite{r14} and
shown to be small (spin splittings $<10\%$), and consequently put
$U_F=1$. The resulting Wilson loop in (\ref{r8}), $W_F\hat
\Phi(x,y)$, consists of two gluon trajectories in $W_F$ and one
straight-line trajectory in $\hat \Phi(x,y)$. The vacuum average
of this Wilson loop produces the area law factor which can be
written as \be \lan W_F\hat \Phi(x,y)\ran = \exp (-\sigma_s (T)
A_{min} (x,y))\label{r11}\ee where $A_{min}(x,y) =\int d^2 \chi
\sqrt{ \det g} $ is the minimal area surface connecting the  three
trajectories discussed above.

For small vacuum correlation length $\lambda, \lambda\ll$ hadron
size,  one can reduce the problem to  that  of a local
Hamiltonian,\cite{r22,r23} \be G^{(2gl)} (x,y) = \lan x
|e^{-H^{(2gl)} t}|y\ran\label{r13}\ee where $H^{(2gl)} $ is
constructed in the same way, as for the $q\bar q$ case in
\cite{r22}, and in the $4d$ case it was written in \cite{r14}.
 One can deduce from (\ref{r13}) the equation for $G^{(2gl)}
 (x,y)$, separating $2d$ space coordinates $\ver_i$ and Euclidean
 time $t$
 \be \left \{ - \frac{\partial}{\partial t} - H^{(2gl)} (\ver_i ,
 \vep_i)\right \} G^{(2gl)} (\ver_i, t;~~ \ver'_i, 0) = \delta
 (t)\prod_{i=1,2} \frac{1}{2\mu_i}
 \delta^{(2)} (\ver_i -\ver'_i)\label{r13a}
 \ee

 In our case of $d=2+1$, the corresponding $2d$ Hamiltonian has
 the form (see Appendix 1 of \cite{r14} for details of derivation).
 \be
 H^{(2gl)} = \frac{\mu_1+\mu_2}{2} +\frac{\vep^2_1}{2\mu_1}
 +\frac{\vep^2_2}{2\mu_2}
 +\sigma_s(T)\{|\ver_1|+|\ver_2|+|\ver_1-\ver_2|\}\label{r14}\ee
 where $\mu_i$ are einbein masses to be found from the
 minimization of eigenvalues \cite{r22}; $\vep_i, \ver_i$ are  $2d$
 vectors of momentum and coordinate of the $i$-th gluon, while the
 position of the adjoint source (the $\hat \Phi(x,y)$ trajectory)
 is chosen at the origin.

 To solve  (\ref{r14}) for the eigenvalues
 $M_n^{(2gl)}(\mu_1,\mu_2)$ it is convenient to use the
 hyperspherical harmonics \cite{r15,r16}, with the global radius $\rho$,
 \be
 \rho^2=\vexi^2+\veta^2,~~ \veta =
 \frac{\ver_1-\ver_2}{\sqrt{2}},~~ \vexi=
 \frac{\ver_1+\ver_2}{\sqrt{2}};\label{r15}\ee
 and the Schroedinger-type  equation for the lowest harmonics
 looks like (one can take $\mu_1=\mu_2=\mu$, as follows from
 minimization anyhow)
 \be
 -\frac{1}{2\mu} \Delta_4 \psi+V(\rho) \psi =E\psi \label{r16}\ee
 with \be \Delta_4 = \frac{d^2}{d\rho^2} \Psi +\frac{3}{\rho}
 \frac{d\Psi}{d\rho} ,~~ V(\rho) = \int d\Omega_4 V(\ver_1,
 \ver_2).\label{r17}\ee

One obtains \be V(\rho) = c_v\rho,~~ c_v \approx \left(1+
\frac{2\sqrt{2}}{3}\right) \sigma_s = \tilde
c\sigma_s\label{r18}\ee A good accuracy (around 1\%) one obtains
when using the simple variational wave function in  (\ref{r16})
\be \Psi(\rho)  = C_0 \exp \left(-\rho^2\frac{\nu^2}{2}\right),~~
E(\mu,\nu) =\frac{c_v}{\nu} \Gamma \left(\frac52\right)
+\frac{\nu^2}{\mu}.\label{r19}\ee

Minimization with respect to the variational parameter $\nu$
yields \be \nu_0 =\left(\frac{\mu
c_v\Gamma\left(5/2\right)}{2}\right)^{1/3},~~ E(\mu,\nu_0)
=\frac32\left(\frac{2}{\mu}\right)^{1/3}
(c_v\Gamma/5/2))^{2/3},\label{r20}\ee
$$ C_0 =\sqrt{2} \nu^2_0, |\Psi(0)|^2=C_0^2 =2\nu^4_0.$$

Finally, the gluelump mass is obtained from minimization over
$\mu$ , $\frac{\partial M_0 (\mu, \nu_0)}{\partial \mu} =
\frac{\partial}{\partial\mu} (\mu + E (\mu, \nu_0))|_{\mu=\mu_0}
=0$ and has \be m_0\equiv M_0 (\mu_0, \nu_0) = 4\mu_0 = 4.52
\sqrt{\sigma_s}.\label{r21}\ee The spectral decomposition of the
gluelump Green's function has the usual form \be G^{(2gl)}
(\mathbf{0}, t; \mathbf{0}  ,0) \equiv
 G^{(2gl)} (t) =\sum^\infty_{n=0} \frac{|\Psi_n (0)
 |^2}{(2\mu_n)^2}
 e^{-M_n^{(2gl)}|t|}\label{r22}\ee
 where $|\Psi_n(0)|^2=\frac{C_n^2}{\Omega_4} =\frac{C^2_n}{2\pi^2}$.
 We  have used in (\ref{r22}) the  standard boson
 normalization $\frac{1}{\sqrt{2\omega}}=\frac{1}{\sqrt{2\mu}}$
 for each gluon,  since $\mu=\mu_0$ is the average energy of the
 gluon, as one can deduce from (\ref{r14}), note that  as in (\ref{r21})  $M_n^{(2gl)} =4 \mu_n$.

 In this way one has from (\ref{r4})
 \be \sigma_s =
\frac{g^4_3(N_c^2-1)}{2} \frac12 \int G^{(2gl)} (x) d^2 x=
\frac{g^4_3(N^2_c-1)}{8\pi} \xi\label{r23}\ee where $\xi$ is \be
\xi=2\pi\int G^{(2gl)} (x) d^2 x = \sum^\infty_{n=0}
\frac{C^2_n}{2\mu_n^2 M_n^2} \equiv \sum^\infty_{n=0}
\xi_n.\label{r24}\ee One can see from (\ref{r20},\ref{r21}) that
the term with $n=0$ contributes to $\xi$ very little
 \be \xi=0.063 +\sum^\infty_{n=1}\xi_n.\label{r25}\ee

The sum over $n$ in (\ref{r24}) is actually diverging, since
$G^{(2gl)} (t)$ as the Green's function of the Hamiltonian
(\ref{r14}), should behave at small $t$ as the free Green's
function, i.e. $G^{(2gl)} (t) \sim t^{-2}$, and the integral $\int
d^2 x G^{(2gl)}(x)$ is logarithmically divergent at $x=0$. This
purely perturbative contribution to $D^H$, is exactly cancelled by
other correlators in the cumulant sum, as shown in \cite{*}. One
can show, in the same way as it was done in \cite{r13} for the 4d
case, that $G^{(2gl)} (t)$ has nonsingular behaviour after
subtraction of the perturbative singularity $O(t^{-2})$. In this
derivation one should take into account that for small loops the
area law behaviour (\ref{r11}) is replaced by the quadratic area
low, $\lan W\ran \sim \exp \left( -\frac{g^2 \lan
F^a_{\mu\nu}\ran^2}{24 N_c} A^2_{min}\right)$, see second ref. in
\cite{r10} for details. To estimate $\xi$, we replace the linear
interaction in $H^{(2gl)}$ by  the oscillator one, (with  $\omega
\approx m_0/2$ to reproduce the asymptotics), and obtain \be \xi
\geq  \int^\infty_{z_0} \frac{zdz}{(\sin  h 2z)^2} , ~~ z_0 \la
\frac{2\omega}{m_0 }=1.\label{25}\ee One can see, that $\xi$ is of
the order of 1, but $\xi < 1$ for reasonable  values of $z_0$.
 Here introduction
of $z_0$ regularizes the perturbative behaviour of $G^{(2gl)}(x)$
at small $x$ and one can see that this interplay of perturbative
and nonperturbative regime at small $x$ of vital importance for
the resulting value of $\xi$, and hence of $\sigma_s$. This should
be compared with lattice data,  where one obtains for $N_c=3$
\cite{r4} \be {\sqrt{\sigma_s}}= g^2_3 0.566= 0.566 T g^2(T);~~
\xi_{lat}\approx 1.00\label{r26}\ee

\section{Glueball excitations in the $3d$ gluodynamics}

As was discussed above, the high-temperature QCD has the $3d$
gluodynamics as a limiting theory, and therefore it is interesting
not only  to  compute field correlators and  $\sigma_s(T)$, as was
done in the previous section, but also to find the spectrum of
this limiting theory.

In this section we shall calculate the spectrum of two-gluon
($2g$) and three-gluon ($3g)$ glueballs in $3d$ SU(N) gluodynamics
and compare results with lattice calculations. The lowest
two-gluon glueball state was previously calculated in the
framework of our formalism in \cite{r12}. We define the creation
operators for glueball states together with spin and parity
$J^{PC}$ in Appendix 1.  One can start from the general QCD string
formalism with spin-dependent terms, as was done in \cite{r17,r18}
in the case of $4d$ QCD, but here we shall neglect both spin
corrections and string rotation corrections \cite{r22}, important
for $L>2$, and consider the simplest version, calculating  with
accuracy of 10\% or  better, so that one has the Hamiltonian for
the $2g$ case \be H=\mu+ \frac{\vep^2}{\mu} + \sigma_a r:~~
H\Psi_n = E_n (\mu) \Psi_n,\label{r27}\ee  with the adjoint string
tension $\sigma_a=\frac{2N^2_c}{N^2_c-1} \sigma_s$, and the
glueball mass is $M_n=E_n(\mu_0)$, where $\mu_0$ is the stationary
point of $E_n(\mu), \frac{\partial
E_n}{\partial\mu}|_{\mu=\mu_0}=0$.  It is known \cite{r24,r25},
that the accuracy of this procedure is around 5-7\%. Instead of
computing (\ref{r27}) numerically, solving  the $2d$ differential
equation, we shall approximate solutions  replacing $\sigma_{a} r
$  by the term $\left( \frac{\sigma^2 r^2}{2\nu}
+\frac{\nu}{2}\right)$ and $\nu$ is later to be found as well as
$\mu$ from the stationary point equation. We shall call the
resulting energy $E_n(\mu,\nu)$ and corresponding masses $\tilde
M_n$. We  expect (and calculations support this) that $\tilde M_n$
differ from $M_n$ within 10\%. As a result one obtains a simple
expression for $\tilde M_n$  \be \tilde M_n=
4\sqrt{\frac{n\sigma_a}{2}},~~ n=1,2,3,...\label{r28}\ee Here
$n=L+1$, and in our approximation radial excited state $0^{+*}$ is
degenerate in mass with the $2^+$ state.

As was mentioned above, we disregard spin-dependent interaction
and therefore the $L=0$ state corresponds to degenerate
$0^{++},2^{++}$ states while $L=2$ state -- to degenerate
$0^{++},2^{++}$, $4^{++}$ states and so on.

The connection (\ref{r28}) is universal for any SU(N) group,
however to connect masses to fundamental spacial string tension
$\sigma_s$ one needs to use the relation \be
\sigma_a=\frac{2N^2_c}{N^2_c-1} \sigma_s,~~ \sigma_a=\frac94
\sigma_s {\rm (SU(3))},~~ \sigma_a=\frac83 \sigma_s {\rm
(SU(2))}~~ \sigma_a= 2\sigma_s{\rm (SU}(\infty))\label{r29}\ee

The resulting ratios $\frac{\tilde M_n}{\sqrt{\sigma_s}}$ are
shown in Table 1  in comparison with lattice data from
\cite{r26}\footnote{The spin-parity assignment in \cite{r26} is in
general different from ours in Appendix 1, and therefore only
states with reliably the same quantum numbers are shown in the
Tables 1 and 2.}. \\


{\bf Table 1}:  The mass ratios $\tilde M_n/\sqrt{\sigma_s}$
obtained from the  Hamiltonian (\ref{r27}) $vs$ lattice data for
angular momentum $L=0,1,2,3,4.$  For comparison  two lowest masses
obtained on the lattice \cite{r26} are given in parentheses.

\begin{center}

\begin{tabular}{|c|c|c|c|c|c|}\hline
&&&&&\\
 $J^{PC}$ & $0^{++},2^{++}$& $0^{-+},2^{-+}$&
$0^{++},2^{++}$&
$0^{-+},2^{-+}$& $0^{++},2^{++}$\\
&&&$4^{++}$&$4^{-+}$&$4^{++},6^{++}$\\&&&&&\\
\hline &&&&&\\L &0&1&2&3&4\\&&&&&\\
\hline &&&&&\\ $\tilde M/\sqrt{\sigma_s}$& 4.61& 6.52& 7.98& 9.22&10.3\\
SU(2)&(4.72)&&(7.82)&&\\
\hline &&&&&\\ $\tilde M/\sqrt{\sigma_s}$ & 4.24&  6&7.34&8.48&9.48\\
SU(3)&(4.33)&&(7.15)&&\\ \hline
&&&&&\\ $\tilde M/\sqrt{\sigma_s}$ & 4& 5.65&6.92&8&8.94\\
SU$(\infty)$&(4.065)&&(6.88)&&\\&&&&&\\ \hline

\end{tabular}

\end{center}

The $3g$ glueballs can be considered in the same way, as it was
done in \cite{r17,r18}
 for the $4d$ case. The Hamiltonian is similar to the one used in
 section 2, Eq. (\ref{r14}) and reads (we put
 $\mu_1=\mu_2=\mu_3=\mu$).
 \be
 H^{(3gl)} = \frac{3\mu}{2} + \frac{\vep^2_\xi+\vep^2_\eta}{2\mu}
 +\sigma_s \sum_{i>j}|\ver_i -\ver_j|.\label{r30}\ee

The hyperspherical component of the grand orbital momentum $K$
satisfies an equation (see Appendix 2 for details of derivation)
\be \left \{ -\frac{1}{2\mu} \left( \frac{d^2}{d\rho^2} +
\frac{2K+3}{\rho} \frac{d}{d\rho}\right) + V(\rho) \right\} \Psi_K
(\rho) =E_K\Psi_K(\rho)\label{r31}\ee with $V(\rho)=c\rho, ~~
c=2\sqrt{2}$.

As is well known (see e.g. \cite{r16}) the solution of (\ref{r30})
is given with 1-2\% accuracy by the equation \be E_{K,n} (\mu) =
W(\rho_0) + \sqrt{\frac{W^{\prime\prime}(\rho_0)}{\mu}} \left( n+
\frac12\right)\label{r32} \ee where $$W(\rho) =\frac{\left(
K+\frac12\right) \left( K+\frac32\right)}{2\mu\rho^2}+ V(\rho),$$
and  $\rho_0$ is the stationary point of $W(\rho),
~W'(\rho)|_{\rho=\rho_0}=0$. The mass $M_{K,n}
=\left\{\frac{3\mu}{2} + E_{K,n}(\mu) \right\}_{min(\mu)} $ is
obtained to be \be M_{K,n} =\sqrt{\sigma_s} 2^{7/4}
(3\Lambda)^{1/4}
\left(1+\frac{2n+1}{\sqrt{3\Lambda}}\right)^{3/4},~~
\Lambda\equiv\left( K+\frac12\right) \left(
K+\frac32\right).\label{r33}\ee The resulting ratio of
$M_{K,n}/\sqrt{\sigma_s}$ are given in the Table 3 for the radial
excitation number $n=0$ and $K=L=0,1,2,3$.\\

{\bf Table 2}:  The  three-gluon glueballs in the $3d$ SU(N)
gluodynamics. Mass ratios $M_{L,0}/\sqrt{\sigma_s}$ are given for
$L=0,1,2,3$. The lowest $C=-1$ mass from lattice data \cite{r26}
is given for comparison in parentheses (see footnote on the
previous page).

\begin{center}

\begin{tabular}{|c|c|c|c|c|}\hline
&&&&\\
K=L &0&1&2&3\\&&&&\\
\hline &&&&\\
$J^{PC}$& $1^{--}, 3^{--}$& $1^{+-},3^{+-}$& $3^{--},1^{--}.5^{--} $& $1^{+-}, 3^{+-}$\\
&&&&\\
\hline &&&&\\
$ \frac{M_{L,0}}{\sqrt{\sigma_s}}$& 6.01& 7.48& 8.68& 9.74\\
&$(5.91, 0^{--})$&&$(8.73, 2^{--})$&\\&&&&\\
\hline
\end{tabular}

\end{center}

Using  Eq. (\ref{r33}) one can also calculate radial excitations,
which are actually collective excitations corresponding to the
collective variable $\rho$. The  resulting mass ratios
$\frac{M_{L,n}}{\sqrt{\sigma_s}}$ for $n=1,2$ are 9.39 and 12.35
respectively.

One can now consider the pomeron and odderon trajectories, using
mass formulas (\ref{r28})  and (\ref{r33}). The pomeron trajectory
corresponds to $J=L+2$, where "2" comes from the sum of "gluon
spins" as in the $d=3+1$ case. In the $d=2+1$ case (as well as for
$d=3+1)$ the creation operators are of the form $tr (D^{L/2} E_i
D^{L/2} E_k$).

From (\ref{r28}) one has \be M^2_J=8\sigma_a (L+1)
=8\sigma_a(J-1)\label{r34a}\ee and the pomeron intercept is
exactly unity, while the Regge slope corresponds to the
relativistic potential case, since string corrections
\cite{r22,r23} are not taken into account.

The odderon trajectory is obtained similarly identifying $J=L+3$,
and finding from (\ref{r33}) (with accuracy better than 1\%)
analytic expression \be M^2(J) =a(J-\Delta_{odd});\label{r35a}\ee
where $a=2^{7/2}\sqrt{3}\sigma_s \cong 19.6 \sigma_s;~~
\Delta_{odd} =2-\frac{\sqrt{3}}{2}  =1.14$.

One can see, that surprisingly the odderon intercept appears
higher, than that of  pomeron, in contrast to the situation found
in the $d=3+1$ QCD \cite{r18}.

\section{Gluerings and gluestars}

For number of gluons $N\geq 3$ one has two options in constructing
gauge-invariant gluon configurations:

a) gluon ring, consisting of $N$ gluons, "sitting" on the closed
loop of fundamental string, with the potential energy $V_{ring} =
\sigma_f \sum_{i=j+1}|r_i-r_j|$;

b) gluon star, consisting of $N$ gluons connected by $N$ adjoint
string pieces with the string junction, where color indices are
assembled in $d^{abc}$ (for $N_c=3$) or in higher order operators
for $N_c>3$. The corresponding potential energy is $V_{star}
=\sigma_a\sum^N_{i=1} |\ver_i -\veY|,$  $\veY$ --  is the string
junction position.

The masses of gluon rings and gluon stars can be computed in the
hyperspherical formalism \cite{r15} using formulas given in
Appendix 2.  One has  from  (\ref{A2.13}) \be
\frac{M_{KN}^{(ring)}}{\sqrt{\sigma_s}}=2(N\Lambda_{KN}
c^2_N)^{1/4}\left( 1+
\frac{2(n+1/2)}{\sqrt{3\Lambda_{KN}}}\right)^{3/4},\label{r34}\ee
with~~~~~ $\Lambda_{KN} =\left( K+N+\frac32\right)\left(
K+N-\frac52\right),$~~~~~~ $c_N = \frac{\sqrt{2}
N\Gamma(N-1)}{\Gamma\left(N-\frac12\right)}\Gamma\left(\frac32\right)$,~~~~~~
and ~~
$\frac{M_{KN}^{(star)}}{M_{KN}^{(ring)}}=\sqrt{\frac{\sigma_a}{\sigma_s}}
\left( \frac{N-1}{2N}\right)^{1/4}$. One can immediately see that
gluestars are 14\% $(N=3)$ to  26\% $(N=\infty)$ heavier than
gluon rings and both saturate at large $N$, i.e.
$\frac{M_{KN}^{(ring)}}{N\sqrt{\sigma_s}} \to  2.24 (N\to
\infty)$.

A similar situation and the same ratios are obtained for $D=3+1$
in \cite{r18}.

\section{Summary and discussion}

An important feature of our results is  that our calculation does
not use any parameters. Several approximations have been made in
obtaining (\ref{r28}), (\ref{r33}): 1)  the use of a local
Hamiltonian, implying small vacuum correlation length $\lambda, ~~
\lambda\ll$ hadron size, and small mixing with hybrids due to the
large mass gap \cite{r27}; 2) spin forces were considered weak and
self-energy corrections vanishing due to renormalization and gauge
invariance (see appendix 4 of \cite{r13} for discussion); 3) color
Coulomb interaction between gluons was neglected due to the
expected strong cancellation with the one-loop correlation,
observed in $4d$ BFKL formalism   (cf. discussion in
\cite{r17,r18}). In $d=4$ glueball calculations \cite{r17,r18}
this suppression of Coulomb resulted in good agreement of
calculated and lattice measured glueball spectrum.

Hence all our approximations are  expected to be accurate within
fifteen percent or  better, which is supported by a good agreement
with lattice data \cite{r26} in Tables 1,2.

On theoretical side the $d=2+1$ gluodynamics was considered also
in \cite{r29}, where the authors developed a Hamiltonian formalism
for the calculation of $\sigma_s$ and vacuum excitations. Their
result for $\sigma_s$ corresponds to $\xi \equiv 1$ in our Eq.
(\ref{r24}) and agrees with lattice data, however the glueball
masses to the knowledge of the author  are not yet available from
the formalism of \cite{r29}.

Summarizing, we have obtained the confining field correlator,
$\sigma_s$, gluelump and glueball masses for $d=2+1$ $SU(N_c)$
gluodynamics. Glueball masses for configurations of different
types: simple glueballs for $N=2$, and gluerings and gluestars for
$N\geq 3$ are computed explicitly without any parameters  through
$\sigma_s$.

Glueball spectrum has a clear hierarchy in growing angular
momentum $L$ and in number of gluons $N$. In this and other
respects the $d=2+1$ spectrum is similar to that of $d=3+1$.

The author is grateful to N.O. Agasian and A.B.Kaidalov for
discussions. The work  is supported
  by the Federal Program of the Russian Ministry of Industry, Science, and Technology
  No.40.052.1.1.1112, and by the
grant for scientific schools NS-1774. 2003. 2.\\

\newpage

\vspace{2cm}

{\it \bf Appendix 1}\\

\begin{center}

{ \large \bf Creation operators of the $d=2+1$ SU(N) glueballs }\\
\end{center}

 \setcounter{equation}{0} \def\theequation{A1.\arabic{equation}}

In $d=2+1$ one can choose $x_3$ as the Euclidean time evolution
coordinate, and denote $E_i=F_{i3}, ~~ H_3=F_{12}$. The gauge
invariant creation operators are constructed from powers of $E_i,
H_3$, and  higher operators also include powers of
$D_i=\partial_i-ig A_i$, with $i=1,2$. The $2d$ angular momentum
$L$ and parity for a given creation operator is defined by the
assignment of $L=1$ to $H_3$ and $D$, $L=0$ to $E_i$. The $2d$
-parity is defined by $PE_i=-E_i, ~PH_3=H_3 , PD_i=-D_i$. The
combination $D_iE_i$ vanishes due equations of motion. The $C$
parity is defined from the transformation $F_{ik}^C \to -
(F_{ik})^T,~~ D_i^C\to -D_i^T$. The resulting operators together
with $L$ and $J^{PC}$ assignments are given in Table 3 for $2g$
glueballs and in Table 4  for  $3g$ glueballs. \\

{\bf Table 3}: Creation operators for $2g$ glueball states
$E_i=F_{i3},~H_3=F_{12}$ in $d=2+1 $ SU(N) theory.

\begin{center}

\begin{tabular}{|c|c|l|}\hline
&&\\
~~~~L ~~~~&$~~~~J^{PC}$~~~~& ~~~~~~Operator\\&&\\
\hline
0&$0^{++}$ &$tr(E_iE_i)$\\
0&$2^{++}$ &$tr(E_iE_k)_{symm.}$\\
1&$2^{-+},0^{-+}$ &$tr(E_iH_3)$\\
2&$2^{++},0^{++}$ &$tr(H_3H_3)$\\
2&$0^{++}$ &$tr(D_iE_kD_iE_k)$\\
2&$2^{++}$ &$tr(D_iE_lD_kE_l)_{symm.}$\\
2&$4^{++}$ &$tr(D_iE_kD_lE_m)_{symm.}$\\
3&$2^{-+}, 0^{-+}$ &$tr(D_iE_kD_iH_3)$\\
3&$4^{-+}, 2^{-+}$ &$tr(D_iE_kD_lH_3)_{symm.}$\\
 4&$2^{++}0^{++}$& $tr(D_iH_3D_iH_3)$\\
 4&$6^{++}, 4^{++}, 2^{++}$& $tr(D_iH_3D_kH_3)_{symm.}$\\ 4&$6^{++}$&
 $tr(D_iD_kE_lD_nD_mE_p)_{symm.}$\\
  \hline

\hline
\end{tabular}

\end{center}

\newpage

{\bf Table 4}: The same as in Table 4,
 but for $3g$ glueballs.

\begin{center}

\begin{tabular}{|c|l|l|}\hline
&&\\
~~~~L ~~~~&$~~~~J^{PC}$~~~~& ~~~~~~Operator\\&&\\
\hline
0&~~~        $1^{--}$ &$tr(E_iE_kE_k)$\\
0&  ~~~      $3^{--}$ &$tr(E_iE_k E_l)_{symm.}$\\
1&    ~~~    $1^{+-}$ &$tr(E_i^2H_3)$\\
1&     ~~~   $3^{+-},1^{+-}$ &$tr((E_iE_k)H_3)_{symm.}$\\
2&    ~~~    $3^{--},1{--}$ &$tr(E_iH_3H_3)$\\
2&   ~~~     $2^{-+},4^{-+}$ &$tr(E_iE_kD_lH_3)$\\
3&   ~~~     $1^{+-},3^{+-}$ &$tr(H_3H_3H_3)$\\
  \hline

\end{tabular}

\end{center}


\vspace{2cm}

{\it \bf Appendix 2}\\

\begin{center}

{ \large \bf Calculation of the $N$--gluon  glueball masses for $N\geq 3$ }\\
\end{center}

 \setcounter{equation}{0} \def\theequation{A2.\arabic{equation}}

The Hamiltonian of $N$-gluon glueballs has the form similar to
(\ref{r30}), we fix the gluon einbein masses to be the same:
$\mu_1=\mu_2=...=\mu$). \be H=\frac{M\mu}{2}
+\frac{\Delta_{2N-2}}{2\mu} + V_N (\ver_1,...
\ver_N)\label{A2.1}\ee where $V_N$ can be of two different forms,
for gluon rings (closed gluon chains with fundamental strings
between adjacent gluons) \be V_N^{(r)} =\sigma_s \sum^{N-1}_{i=0}
|\ver_{i+1} -\ver_i|,~~ \ver_N \equiv \ver_0\label{A2.2}\ee and
for gluon stars ( a generalization of the $Y$-type glueball for
$N_c=3$ with the color symmetric matrix $d^{abc}$ at the string
junction vertex, see \cite{r18} for details) \be V_N^{(star)}
=\sigma_a \sum^N_{i=1} |\ver_i-\veY|\label{A2.3}\ee where $\veY$
is the position of the string junction, which for gluons can be
taken to coincide with the c.m. coordinate $\veR =\frac{1}{N}
\sum^N_{i=1} \ver_i$.

We approximate the eigenfunction of $H$ (\ref{A2.1}) for a given
angular momentum $L$ by the hyperspherical component \cite{ } \be
\Psi_K (\ver_1, \ver_2,...\ver_N)=
\frac{\chi_K(\rho)}{\rho^{2K+N+\frac32}} \mathcal{P}_K (\ver_1,
\ver_2,... \ver_N)\label{A2.4}\ee where $\rho^2=\sum^N_{i=1}
(\ver_i-\veR)^2$, and  $ \mathcal{P}_K$ is the harmonic polynomial
of Jacobi coordinates $\vexi_i , i=1, N-1$, constructed of
$\ver_1,...\ver_N$, \be \vexi_i =\frac{1}{\sqrt{i(i+1)}} \left[
\sum^i_{j=1} \ver_j -i\ver_{i+1}\right]\label{A2.5}\ee

Averaging of $(\Psi_K H\Psi_K)$ over angular variables
$\Omega_{2N-2}$ in $(2N-2)$ space  of $\{\vexi_i\}$ yields an
equation for $ \chi_K(\rho)$ \be \left( - \frac{1}{2\mu}
\frac{d^2}{d\rho^2} +W_{KN} (\rho) +\frac{N\mu}{2} \right)\chi_K
(\rho)=M_{KN} (\mu) \chi_K (\rho)\label{A2.6}\ee where $W_{KN}
(\rho)$ is \be W_{KN} (\rho) =\frac{\Lambda_{KN}}{2\mu\rho^2} +
V^{(k)}_N (\rho),~~ \Lambda_{KN} \equiv \left(
K+N-\frac32\right)\left( K+N-\frac52\right)\label{A2.7}\ee and \be
V_N^{(k)} (\rho) \equiv \lan V_N^{(k)} (\ver_1,
\ver_2,...)\ran_\Omega,~~ k=r, ~{\rm star}.\label{A2.8}\ee

To calculate $V_N^{(k)} (\rho)$ it is  enough to know the angular
average of one Jacobi vector, \be V_N^{(k)} (\rho)= \sigma_s
N\lambda^{(k)} \lan |\vexi_i|\ran_\Omega=\sigma_s \rho N
\frac{\lambda^{(k)}\Gamma(N-1) \Gamma
\left(\frac32\right)}{\Gamma\left(N-\frac12\right)}\equiv
C_N^{(k)}\rho\sigma_s\label{A2.9}\ee and $\lambda^{(k)} =\sqrt{2}$
for $k=r$(ring), and $\lambda^{(star)}
=\sqrt{\frac{N-1}{N}}\frac{\sigma_a}{\sigma_s}$, as one can deduce
from (\ref{A2.5}) for $i=N-1$.

Finally the solution for $M_{Kn} (\mu)$ can be found with $\sim
1\%$ accuracy from the minimum of $W_{KN} (\rho)$ as a function of
$\rho$ \cite{r15,r16,r28}. \be \frac{m_{KN}}{\sqrt{\sigma_s}}
(\mu)= W_{KN} (\rho_0) +\omega \left( n+\frac12\right),~~
\omega^2=\frac{1}{\mu} W^{\prime\prime}(\rho_0)\label{A2.10}\ee
where $\rho_0 =\left( \frac{\Lambda_{KN}}{\mu
c_N^{(k)}}\right)^{1/3}$, so that one has \be
\frac{m_{KN}(\mu)}{\sqrt{\sigma_s}} = \frac{\eta_N}{\mu^{1/3}},~~
\eta_N =\frac32 \Lambda_{KN}^{1/3} (c_n^{(k)})^{2/3} \left(
1+\frac{2(n+1/2)}{\sqrt{3\Lambda_{KN}}}\right).\label{A2.11}\ee

From $\frac{\partial M_{KN}(\mu)}{\partial \mu} |_{\mu=\mu_0}=0$
one has \be
\frac{M_{KN}(\mu)}{\sqrt{\sigma_s}}=\frac{N\mu}{2}+\frac{\eta_N}{\mu^{1/3}},~~
\mu_0=\left(\frac{2\eta_N}{3N}\right)^{3/4}\label{A2.12}\ee and
finally  the mass of the $N$ gluon state is \be
\frac{M_{KN}}{\sqrt{\sigma_s}} = 2 B_N^{1/4} \left( 1+
\frac{2\left(
n+\frac12\right)}{\sqrt{3\Lambda_{KN}}}\right)^{3/4}\label{A2.13}\ee
where
$$
B_N=N \Lambda_{KN} (C^{(k)}_N)^2.$$

For $N=3$ and $K=0, B_3=18$ and
$\frac{M_{03}(n)}{\sqrt{\sigma_s}}= 6.01, 9.39, 12.35$ for
$n=0,1,2$ respectively.

For $N$ large, one can consider a gluon ring of $N$ gluons
connected by fundamental string for any $N_c$.

Then asymptotically $C_{K,N} (N\to \infty) \approx
\sqrt{\frac{\pi}{2}} \frac{N}{\sqrt{N-1}},$  $ B^{1/4}_{N\to
\infty} = N\left(\frac{\pi}{2}\right)^{1/4}$ and \be \frac{M_{KN}
(N\to\infty)}{N\sqrt{\sigma_s}} =
2\left(\frac{\pi}{2}\right)^{1/4} \approx 2.24.\label{A2.14}\ee

For $N=3$ this asymptotic formula yields
$\frac{M}{\sqrt{\sigma_s}}\approx 6.72$ which is not far from our
exact value of 6.01.

The gluon stars are obtained with $\lambda^{(star)} $ in
$C_N^{(k)}$ instead of $\lambda^{(ring)}$, which yields for the
mass the relative factor
$\sqrt{\frac{\lambda^{(star)}}{\lambda^{(ring)}}}=
\sqrt{\frac{\sigma_a}{\sigma_s}}\cdot \left(
\frac{N-1}{2N}\right)^{1/4}$. For $N=3$ this factor is 1.14, which
means, that  the 3-gluon star is 14\% heavier than  the 3-gluon
ring.

 \end{document}